\documentclass[a4paper,11pt]{article}
\pdfoutput=1

\usepackage[a4paper,left=85pt,right=75pt,top=80pt,bottom=100pt]{geometry}
\usepackage[latin1]{inputenc}
\usepackage{dsfont}
\usepackage{amsfonts,amsmath,amssymb}
\usepackage{amsthm}
\usepackage{graphicx}
\usepackage[small,bf]{caption}
\usepackage{subcaption}
\usepackage{booktabs}
\usepackage[numbers, sort]{natbib}
\allowdisplaybreaks[1]
\usepackage{color}
\usepackage{ulem}
\usepackage{tikz}
\usepackage{multirow}
\usetikzlibrary{arrows}

\usepackage[pagebackref=false]{hyperref}

\def\bal#1\eal{\begin{align}#1\end{align}}
\newcommand{\be}{\begin{equation}}
\newcommand{\ee}{\end{equation}}
\newcommand{\besub}{\begin{subequations}}
\newcommand{\eesub}{\end{subequations}}
\newcommand{\ba}{\begin{array}}
\newcommand{\ea}{\end{array}}
\newcommand{\bi}{\begin{itemize}}
\newcommand{\ei}{\end{itemize}}

\newcommand{\TeV}{{\rm TeV}}

\newcommand{\ckm}{{\text{\tiny CKM}}}
\newcommand{\mns}{{\text{\tiny PMNS}}}
\newcommand{\sm}{{\text{\tiny SM}}}
\newcommand{\mssm}{{\text{\tiny MSSM}}}
\newcommand{\GUT}{{\text{\tiny GUT}}}
\newcommand{\SUSY}{{\text{\tiny SUSY}}}
\newcommand{\MSbar}{{\overline{\text{MS}}}}
\newcommand{\DRbar}{{\overline{\text{DR}}}}

\def\diag{\mathop{\rm diag}}

\begin{document}

\begin{titlepage}

\begin{center}
{\LARGE
Running quark and lepton parameters at various scales
}
\\[15mm]

Stefan Antusch$^{\star\dagger}$\footnote{Email: \texttt{stefan.antusch@unibas.ch}},~
Vinzenz Maurer$^{\star}$\footnote{Email: \texttt{vinzenz.maurer@unibas.ch}},~
\end{center}
\addtocounter{footnote}{-2}

\vspace*{0.20cm}

\centerline{$^{\star}$ \it
Department of Physics, University of Basel,}
\centerline{\it
Klingelbergstr.\ 82, CH-4056 Basel, Switzerland}

\vspace*{0.4cm}

\centerline{$^{\dagger}$ \it
Max-Planck-Institut f\"ur Physik (Werner-Heisenberg-Institut),}
\centerline{\it
F\"ohringer Ring 6, D-80805 M\"unchen, Germany}

\vspace*{1.2cm}

\begin{abstract}

We calculate the values and $1\sigma$ ranges of the running quark and lepton Yukawa couplings as well as of the quark mixing parameters at various energy scales, 
i.e.\ at $M_Z$, 1 TeV, 3 TeV, 10 TeV and at the GUT scale, to provide useful input for model building. Above TeV energies, we assume the minimal supersymmetric extension of the Standard Model, and include the $\tan \beta$-enhanced one-loop supersymmetric threshold corrections which arise when matching the SM to its SUSY extension. We calculate the GUT scale values of the running parameters as well as their $1\sigma$ ranges, with the supersymmetric threshold corrections included and with their effects parametrised in a simple way.

\end{abstract}

\end{titlepage}
\newpage

\section{Introduction}

Our knowledge about the fermion masses and mixing parameters has improved significantly in the last years. For instance in the lepton sector, the mixing angle $\theta_{13}^\mns$ of the Pontecorvo-Maki-Nakagawa-Sakata (PMNS) matrix has recently been measured \cite{t13_data}, and the precision for the other leptonic mixing angles has increased \cite{GonzalezGarcia:2012sz}. In parallel, also the results for light quark masses have improved significantly, as reported recently by the Flavour Averaging Group (FLAG) \cite{flag} and as stated in the recent update of the Particle Data Group (PDG) review \cite{pdg}. The combined results have become powerful constraints on models for fermion masses and mixing.

To construct models, it is very useful \cite{Xing:2007fb} to have the running fermion masses (or Yukawa couplings) and mixing parameters at hand at the (high) energy scale where the model is defined, or, alternatively, at a common low energy scale (e.g.\ at the scale $M_Z$ of the Z boson mass) where it is convenient to compare the model predictions with the experimental data after their renormalisation group (RG) evolution from high energy to $M_Z$ has been performed. In the former case, an example is provided by Grand Unified Theories (GUTs) where the model is defined at $M_{\GUT} \simeq 2 \times 10^{16}$ GeV. GUTs are often formulated in the framework of supersymmetry (SUSY), which implies that the supersymmetric radiative threshold corrections \cite{SUSYthresholds,Blazek,Antusch:2008tf,Crivellin:2012zz} have to be included when deriving the running parameters at $M_{\GUT}$. 

In this paper, we calculate the values and allowed ranges of the running quark and charged lepton parameters at various energy scales from the experimental results provided at low energies, to provide useful input for model building. Above TeV energies, we assume the minimal supersymmetric extension of the Standard Model (MSSM), and include the $\tan \beta$-enhanced one-loop supersymmetric threshold corrections. Their effects are parametrised in a simple way, such that they can readily be included in model building considerations.

The paper is organised as follows: In section 2 we describe our numerical procedure and specify the used low energy input values. Section 3 contains our parametrisation for including the SUSY threshold corrections. In sections 4 and 5 we present our results for the parameters at $\mu = M_Z$, $1$ TeV, $3$ TeV and $10$ TeV and at $M_\GUT$, respectively. In section 6 we give a summary.

\section{Numerical analysis}\label{numerics}

Using the same notation as in \cite{Chetyrkin:2000yt}, with $M_i$ denoting pole masses and $m_i(\mu)$ the running masses, the low energy input values used in our analysis are as follows: 

For the light quark masses (at $\mu = 2$ GeV, with $n_f=2+1$ active flavours), we use the values from the latest PDG review \cite{pdg}, 
$m_s/m_{ud} = 27 \pm 1$ (with $m_{ud} = (m_u + m_d)/2$),
$m_d = 4.8^{+0.7}_{-0.3}$ MeV, and
$m_s = 95 \pm 5$ MeV.
Furthermore, also from \cite{pdg}, we use
$m_c(m_c) = 1.275 \pm 0.025$ GeV ($n_f=4$), 
$m_b(m_b) = 4.18 \pm 0.03$ GeV ($n_f=5$),
$M_t = 173.15 \pm 1.4$ GeV (pole mass),
$\alpha_s(M_Z) = 0.1184 \pm 0.0007$ (with $n_f=5$ active flavours),
$M_e = 0.510998928 \pm 0.000000011$ MeV,
$M_\mu = 105.6583715 \pm 0.0000035$ MeV,
$M_\tau = 1776.82 \pm 0.16$ MeV,
$1/\alpha(M_Z)  = 127.944 \pm 0.014$, and
$\hat{s}^2_{\theta_W} = 0.23116 \pm 0.00012$. 
For the quark mixing parameters, we use the values from the UTFit collaboration \cite{UTfit}: 
$\sin\theta_{12} = 0.2254 \pm 0.0007$,
$\sin\theta_{23} = 0.04207 \pm 0.00064$,
$\sin\theta_{13} = 0.00364 \pm 0.00013$, and 
$\delta = 1.208 \pm 0.054$. The Higgs self-coupling is extracted from its measured mass $m_h = 125.8 \pm 0.8$ from \cite{CMSHiggs}. 

The input values for the QCD parameters (quark masses and $\alpha_s$) are evolved to $M_Z = 91.1876$ GeV~\cite{pdg} using the Mathematica package RunDec \cite{Chetyrkin:2000yt}, integrating in the top quark at $M_Z$ as described therein. The running lepton masses at $M_Z$ are computed from their pole masses as in \cite{Arason:1991ic}. The masses are then converted to Yukawa couplings according to $m_f = y_f v$ with the Higgs VEV $v = 174.104$ GeV as obtained from $G_F$, while $\alpha_s$, $\alpha$ and $\sin\theta_W$ are used to calculate the gauge couplings $g_i$. 

The parameters at $M_Z$ are passed to a modified version of the Mathematica package REAP \cite{Antusch:2005gp}, which is calculating the running to the desired higher scales. The modifications introduced to the current version of REAP include two loop SM and MSSM RGEs \cite{Luo:2002ey,Martin:1993zk} for all quantities, automatic conversion of $\MSbar$ to $\DRbar$ quantities \cite{Martin:1993yx} at the SM-MSSM threshold as well as the handling of $\tan\beta$ enhanced threshold corrections as described in the next section.\footnote{We will assume that neutrino masses are generated via the seesaw mechanism at high energies, and that we are in a regime with small neutrino Yukawa couplings such that their effects on the RG evolution can be neglected. For obtaining the neutrino masses and leptonic mixing parameters at the GUT scale, which depend on various additional parameters, we recommend to solve the RGEs for a given model directly with REAP or to use as approximations the analytical formulae presented in \cite{Antusch:2005gp}.}

\section{Inclusion of the SUSY threshold corrections}\label{sec:thrcor}
For calculating parameters at high energies in a SUSY theory, the SM has to be matched to its SUSY extension. The common approximation, which we will also use in this study, is to do this at a single threshold scale $M_\SUSY$ where all superpartners are integrated out at once. In particular for moderate or large $\tan \beta$, the radiative threshold corrections \cite{SUSYthresholds,Blazek,Antusch:2008tf,Crivellin:2012zz} can be large since some of the contributing diagrams are $\tan \beta$ enhanced. Due to this enhancement, they can even exceed the one-loop running contribution. It is therefore mandatory to include them when analysing the running quark and charged lepton masses in SUSY models.  

In our calculation of the parameters at the GUT scale, we will include the $\tan \beta$ enhanced 1-loop threshold effects. They result in a correction to the Yukawa matrices of the down-type quarks and charged leptons at the matching scale between the SM and the MSSM. 
This correction is most conveniently described in the basis where the up-type quark Yukawa matrix $Y_u$ is diagonal. Then one can write the matching conditions with some simplifying approximations as \cite{Blazek} 
\begin{align}
\label{eq1}    Y_u^\sm &\simeq Y_u^\mssm \sin\beta \:, \\
    Y_d^\sm &\simeq (\mathds{1} + \diag(\eta_q, \eta_q, \eta'_q + \eta_A)) \, Y_d^\mssm \cos\beta \:,\\
\label{eq3}     Y_e^\sm &\simeq (\mathds{1} + \diag(\eta_\ell , \eta_\ell, \eta'_\ell )) \,Y_e^\mssm \cos\beta \:.
\end{align}
We drop here contributions which are not $\tan \beta$ enhanced. They are expected to give further corrections at less than percent-level. Neglecting them is consistent since we will not claim better accuracy for the GUT scale masses (respectively Yukawa couplings). 
We also approximate that the first two generations of down-type quarks and charged leptons each receive the same threshold corrections, which is a good approximation as long as their superpartners, i.e. the down squark and the strange squark respectively the selectron and the smuon have nearly the same mass, which is a common feature of many SUSY scenarios. 

Eqs.~(\ref{eq1}) - (\ref{eq3}) contain six parameters: $\eta_q$ and $\eta'_q$ are dominated by the gluino contribution but also receive corrections from loops with Winos and Binos.  The parameters $\eta_\ell$ and $\eta'_\ell$ only stem from electro-weak gauginos and are thus often smaller than $\eta_q$ and $\eta'_q$. $\eta_A$ is a correction due to the trilinear soft SUSY breaking term $A_u$ (cf.\ \cite{Blazek}) which we assume to be hierarchical (as the quark and charged lepton Yukawa matrices). An analogous correction does not appear for the charged leptons due to the absence of light right-handed neutrinos.
All the $\eta$'s contain a factor of $\tan \beta$, and are often written as $\eta_i = \varepsilon_i \tan \beta$. They can be calculated explicitly once a SUSY scenario is specified. Formulae for the 1-loop results in the electroweak-unbroken phase can be found e.g.\ in \cite{SUSYthresholds,Antusch:2008tf}. 

From Eqs.~(\ref{eq1}) - (\ref{eq3}) one can see that the Yukawa matrices only depend on four combinations of the six parameters, while two can be absorbed: The parameter $\eta'_\ell$ can be absorbed in a re-definition of $\beta \to \bar \beta$, such that
\be\label{eq:def_barbeta}
\cos\bar{\beta} := (1 + \eta'_\ell) \cos\beta \:,
\ee
or equivalently (to a good approximation) $\tan\bar{\beta} := (1 + \eta'_\ell)^{-1} \tan\beta$. Introducing furthermore the parameters $\bar{\eta}_b$ , $\bar{\eta}_q$ and  $\bar{\eta}_\ell$ as
\begin{align}
\label{eq:def_baretaB}  \bar{\eta}_b   &:=   \eta'_q + \eta_A - \eta'_\ell \:,\\
  \bar{\eta}_q  &:= \eta_q - \eta'_\ell\:,\\
   \bar{\eta}_\ell  &:= \eta_\ell - \eta'_\ell\:,
\end{align}
we can rewrite the matching conditions of Eqs.~(\ref{eq1}) - (\ref{eq3}) as
\begin{align}
 \label{eq:matching_u}   Y_u^\sm &\simeq Y_u^\mssm \sin\bar{\beta} \:,\\
 \label{eq:matching_d}   Y_d^\sm &\simeq (\mathds{1} + \diag(\bar{\eta}_q, \bar{\eta}_q, \bar{\eta}_b)) \,Y_d^\mssm \cos\bar{\beta} \:,\\
 \label{eq:matching_e}   Y_e^\sm &\simeq  (\mathds{1} + \diag(\bar\eta_\ell ,\bar \eta_\ell, 1 )) \,Y_e^\mssm \cos\bar{\beta} \:.
\end{align}
We consider moderate or large $\tan \beta$ here (i.e.\ $\tan \beta \ge 5$), where we can approximate $\sin\beta \simeq \sin\bar{\beta}$. We remark that, as mentioned above, the lepton correction parameters $\eta_\ell,\eta'_\ell$ are typically smaller than $\eta_q,\eta'_q$ and are therefore often neglected. In our parametrisation neglecting $\eta_\ell,\eta'_\ell$ would simply mean that $\bar \beta = \beta$. However, since the effects of $\eta_\ell$ and $\eta'_\ell$ can be relevant (cf.\ e.g.\ \cite{Antusch:2008tf}), we prefer to include them in the analysis. 

In table \ref{tab:SMDRbarvalues} we will give the values of the running SM parameters, converted to the $\DRbar$ scheme used in the analysis above $M_\SUSY$. They can be used to calculate the MSSM Yukawa matrices $Y_u$, $Y_d$ and $Y_e$ at $M_\SUSY$, with threshold corrections included, from  Eqs.~(\ref{eq:matching_u}) - (\ref{eq:matching_e}). Explicitly, in the basis where $Y_u$ and $Y_e$ are diagonal we obtain the Yukawa matrices from table \ref{tab:SMDRbarvalues} as (using PDG parametrisation for the Yukawa matrices)
\begin{align}
Y_u^\sm &= \diag(y^\sm_u, y^\sm_c, y^\sm_t) \:,\\
Y_d^\sm &= V_\ckm^\dagger (\theta^{q,\sm}_{12},\theta^{q,\sm}_{23},\theta^{q,\sm}_{13},\delta^{q,\sm}) \,\diag(y^\sm_d, y^\sm_s, y^\sm_b)\:,\\
Y_e^\sm &= \diag(y^\sm_e, y^\sm_\mu, y^\sm_\tau) \:.
\end{align} 
With these expressions for the SM Yukawa matrices, the MSSM Yukawa matrices at $M_\SUSY$ are then given as
\begin{align}
 \label{eq:matching_u_2}Y_u^\mssm &\simeq  Y_u^\sm \,\frac{1}{\sin\bar{\beta}} \:,\\
 \label{eq:matching_d_2}Y_d^\mssm &\simeq  \diag \left(\frac{1}{1+\bar{\eta}_q}, \frac{1}{1+\bar{\eta}_q},\frac{1}{1+ \bar{\eta}_b}\right)\,Y_d^\sm\,\frac{1}{\cos\bar{\beta}}\:,\\
 \label{eq:matching_e_2}Y_e^\mssm &\simeq  \diag \left(\frac{1}{1+\bar{\eta}_\ell}, \frac{1}{1+\bar{\eta}_\ell},1\right)\,Y_e^\sm\,\frac{1}{\cos\bar{\beta}} \:.
\end{align} 

Finally, we note that the parameters $\bar{\eta}_q$ and $\bar{\eta}_\ell$ only correct the Yukawa couplings of the first two generations of down-type quarks and charged leptons, whereas $\bar{\eta}_b$ affects the third generation. This has the following advantage: Since $\bar{\eta}_q$ and $\bar{\eta}_\ell$ only induce a correction to Yukawa couplings which are comparatively small, their effect can be neglected in the beta-functions when calculating the RG evolution of the parameters. To a good approximation, the RG evolution thus only depends on $\bar{\eta}_b$ and $\tan\bar{\beta}$. This will be useful to simplify the discussion of the effects of the SUSY threshold corrections on the running parameters at $M_{\mathrm{GUT}}$ in section~\ref{results}.

\section{Results at $\boldsymbol{\mu = M_Z}$, 1 TeV, 3 TeV and 10 TeV}\label{results}

The results at $\mu = M_Z$, $1$ TeV, $3$ TeV and $10$ TeV, calculated within the SM in the $\MSbar$ scheme, are given in table \ref{tab:SMvalues}. The values converted to the $\DRbar$ scheme can be found in table \ref{tab:SMDRbarvalues}. To calculate the uncertainties, we took the uncertainties of the input parameters as stated in section \ref{numerics} and performed a Monte Carlo analysis to obtain the (marginalised) highest posterior density (HPD) intervals (corresponding to the $1\sigma$ uncertainties) for the parameters at the desired scale.

\section{Results at the GUT scale}\label{resultsGUT}

To obtain the running parameters at the GUT scale (using $M_\GUT = 2 \times 10^{16}$ GeV) we perform the matching from the SM to the MSSM at the scale $M_\SUSY$, as described in section \ref{sec:thrcor}. 

From Eqs.~(\ref{eq:matching_u_2}) - (\ref{eq:matching_e_2}), in leading order in small mixing approximation, one can explicitly obtain the relations between the eigenvalues of the SM and MSSM Yukawa matrices as
\begin{align}
\label{eq:thres_yu}   y^{\sm}_{u,c,t} &\simeq y^{\mssm}_{u,c,t} \sin \bar\beta \;,\\
\label{eq:thres_yd}    y^{\sm}_{d,s} &\simeq (1 + \bar\eta_q)~ y^{\mssm}_{d,s} \cos  \bar\beta \;,\\
\label{eq:thres_yb}    y^{\sm}_{b} &\simeq (1 + \bar\eta_b)~ y^{\mssm}_b \cos  \bar\beta \;,\\
\label{eq:thres_ye}    y^{\sm}_{e,\mu} &\simeq (1 +\bar\eta_\ell)~ y^{\mssm}_{e,\mu} \cos \bar\beta \;,\\
\label{eq:thres_ytau}    y^{\sm}_{\tau} &\simeq y^{\mssm}_{\tau} \cos \bar\beta \;.
\end{align}
The relations between the SM and MSSM mixing parameters of the CKM matrix are (again in leading order in a small mixing approximation) 
\begin{align}
\label{eq:thres_ti3}    \theta^{q,\sm}_{i3} &\simeq \frac{1 + \bar\eta_q}{1 + \bar\eta_b}~ \theta^{q,\mssm}_{i3} \;, \\
\label{eq:thres_t12}   \theta^{q,\sm}_{12} &\simeq \theta^{q,\mssm}_{12} \;, \\
\label{eq:thres_delta}    \delta^{q,\sm} &\simeq \delta^{q,\mssm} \;.
\end{align}
To a very good approximation, $\theta^{q,\sm}_{12}$ and $\delta^{q,\sm}$ are not affected by the threshold corrections and are also stable under the RG evolution. After having calculated the MSSM quantities, their running between $M_\SUSY$ and $M_\GUT$ depends to a good approximation only on $\bar\eta_b$ and $\tan\bar\beta$, as discussed in section 3. This allows to present the GUT scale values of the quantities on the right sides of the equations as functions of $\bar\eta_b$ and $\tan\bar\beta$ only (for a fixed $M_\SUSY$).  

Our results for the GUT scale quantities (assuming $M_\SUSY = 1$ TeV) are shown in figures 1 to 5, and the relative 1$\sigma$ uncertainties for the GUT scale parameters are given in table 3. For the third generation Yukawa couplings, the allowed ranges depend on $\bar\eta_b$ and $\tan\bar\beta$, and are given in figure \ref{fig:GUTscalevaluesUncertainty}. The relative uncertainties for the other parameters are practically independent of $\bar\eta_b$ and $\tan\bar\beta$ for most of the parameter space. Only in the parameter regions in figure 1, 2, 3 and 5 very close to the grey area, where one of the third generation Yukawa couplings gets non-perturbatively large, the uncertainties also increase accordingly. In addition, we also show the values of the GUT scale ratios $y_t/y_b$ and $y_\tau/y_b$ as functions of $\tan\bar\beta$ and $\bar{\eta}_b$ in figure \ref{fig:GUTscalevaluesRatios}.
The GUT scale ratios $y_\mu/y_s$ and $y_e/y_d$ are insensitive to $\tan\bar\beta$ and $\bar{\eta}_b$ to a good approximation, since the factor $\tan\bar\beta$ cancels out in the matching conditions and since $\bar{\eta}_b$ affects the RG evolution of both quantities in the same way (via the trace terms in the RGEs). The ratios at $M_\GUT$ (assuming $M_\SUSY = 1$ TeV) can be obtained from the following formulae:
 \be
 \frac{(1 + \bar\eta_\ell) y_\mu}{(1 + \bar\eta_q) y_s} \approx 4.36 \pm 0.23 \; , \quad 
 \frac{(1 + \bar\eta_\ell) y_e}{(1 + \bar\eta_q) y_d} \approx 0.41^{+0.02}_{-0.06} \; .
 \ee
One can also derive the following relation at $M_\GUT$, 
 \be\label{eq:4ys}
 \frac{y_\mu}{y_s} \frac{y_d}{y_e} \approx 10.7^{+1.8}_{-0.8} \: ,
 \ee
where the dependence on the threshold correction parameters has dropped out. Furthermore, the relation in Eq.~(\ref{eq:4ys}) is also not affected by a change of $M_\SUSY$. 

Let us briefly comment on the dependence of the other GUT scale quantities on $M_\SUSY$. Generally speaking, changing $M_\SUSY$ from 1 TeV to 10 TeV can lead to changes of the GUT scale quantities at the few \%-level, which furthermore also depend on $\tan\bar\beta$ and $\bar{\eta}_b$. 
We provide our results also in the form of data tables, available at \url{http://particlesandcosmology.unibas.ch/RunningParameters.tar.gz}, which in addition to $M_\SUSY = 1$ TeV also include the results for $M_\SUSY = 3$ TeV and $10$ TeV. From these tables, using a numerical interpolation (cf.\ example Mathematica notebook provided in the above mentioned RunningParameters.tar.gz file), one can easily obtain the GUT scale quantities as functions of the relevant parameters such that they can be used conveniently for model analyses.

\section{Summary and conclusions}

In this paper, we have calculated the values of the running quark and lepton masses as well as of the quark mixing parameters at various energy scales, to provide useful input for model building. The results at $\mu = M_Z$, $1$ TeV, $3$ TeV and $10$ TeV, calculated within the SM in the $\MSbar$ scheme, are given in table \ref{tab:SMvalues}. The values converted to the $\DRbar$ scheme can be found in table \ref{tab:SMDRbarvalues}. They can be used as convenient input for fitting a model after having performed the RG evolution of the parameters to the respective scale.

In the second part of the analysis, we calculated the GUT scale values of the running parameters in the $\DRbar$ scheme, assuming the MSSM above $M_\SUSY = 1$ TeV. We present our results using a novel parametrisation for including the $\tan \beta$-enhanced one-loop SUSY threshold corrections. The GUT scale parameters can be obtained in a simple way as functions of $\bar{\eta}_b$ and $ \tan\bar{\beta}$, with $\bar\beta$ and $\bar{\eta}_b$ defined in Eqs.~(\ref{eq:def_barbeta}) and (\ref{eq:def_baretaB}), respectively. 

We furthermore provide the numerical results for the running parameters at $M_\GUT$ in figures $1$ to $5$ and table 3, and also in form of data tables (at \url{http://particlesandcosmology.unibas.ch/RunningParameters.tar.gz}). The latter allow to reproduce the GUT scale quantities as numerical functions for convenient use in model analyses. In addition to the case $M_\SUSY = 1$ TeV, the data tables also contain the results for $M_\SUSY = 3$ TeV and $10$ TeV. 
With these results, the fit of GUT scale models to the experimental data can be greatly simplified and accelerated.

\section*{Acknowledgements}
This project was supported by the Swiss National Science Foundation (projects 200021-137513 and CRSII2-141939-1).


\newpage

\begin{table}
\hspace{-0.7cm}
\setlength{\tabcolsep}{0.3em}
\begin{tabular}{|c|ll|ll|ll|ll|}
\hline
\rule{0 px}{11 px}
SM Quantity & \multicolumn{2}{|c|}{$\mu = M_Z$} & \multicolumn{2}{|c|}{$= 1$ TeV} & \multicolumn{2}{|c|}{$= 3$ TeV} & \multicolumn{2}{|c|}{$= 10$ TeV} \\
\hline
\hline
\rule{0 px}{13 px}
$y^\sm_u \;/\, 10^{-6}$ & $7.4$ & $^{+1.5}_{-3.0}$ & $6.3$ & $^{+1.3}_{-2.6}$ & $6.0$ & $^{+1.2}_{-2.5}$ & $5.7$ & $^{+1.2}_{-2.3}$ \\
\rule{0 px}{15 px}
$y^\sm_d \;/\, 10^{-5}$ & $1.58$ & $^{+0.23}_{-0.10}$ & $1.364$ & $^{+0.198}_{-0.087}$ & $1.291$ & $^{+0.188}_{-0.082}$ & $1.223$ & $^{+0.180}_{-0.076}$ \\
\rule{0 px}{15 px}
$y^\sm_s \;/\, 10^{-4}$ & $3.12$ & $^{+0.17}_{-0.16}$ & $2.70$ & $^{+0.14}_{-0.15}$ & $2.56$ & $^{+0.13}_{-0.14}$ & $2.42$ & $^{+0.12}_{-0.13}$ \\
\rule{0 px}{15 px}
$y^\sm_c \;/\, 10^{-3}$ & $3.60$ & $\pm 0.11$ & $3.104$ & $\pm 0.095$ & $2.935$ & $^{+0.090}_{-0.091}$ & $2.776$ & $^{+0.084}_{-0.088}$ \\
\rule{0 px}{15 px}
$y^\sm_b \;/\, 10^{-2}$ & $1.639$ & $\pm 0.015$ & $1.388$ & $^{+0.013}_{-0.014}$ & $1.303$ & $\pm 0.013$ & $1.224$ & $^{+0.013}_{-0.012}$ \\
\rule{0 px}{15 px}
$y^\sm_t$ & $0.9861$ & $^{+0.0086}_{-0.0087}$ & $0.8685$ & $^{+0.0090}_{-0.0084}$ & $0.8278$ & $^{+0.0089}_{-0.0085}$ & $0.7894$ & $^{+0.0083}_{-0.0092}$\\[0.5ex]
\hline
\rule{0 px}{15 px}
$\theta_{12}^{q,\sm}$ & $0.22735$ & $\pm 0.00072$ & $0.22736$ & $^{+0.00072}_{-0.00071}$ & $0.22736$ & $^{+0.00072}_{-0.00071}$ & $0.22736$ & $^{+0.00072}_{-0.00071}$ \\
\rule{0 px}{15 px}
$\theta_{23}^{q,\sm} \;/\, 10^{-2}$ & $4.208$ & $\pm 0.064$ & $4.296$ & $^{+0.066}_{-0.065}$ & $4.330$ & $\pm 0.066$ & $4.364$ & $^{+0.067}_{-0.066}$ \\
\rule{0 px}{15 px}
$\theta_{13}^{q,\sm} \;/\, 10^{-3}$ & $3.64$ & $\pm 0.13$ & $3.72$ & $\pm 0.13$ & $3.75$ & $\pm 0.13$ & $3.77$ & $^{+0.13}_{-0.14}$ \\
\rule{0 px}{15 px}
$\delta^{q,\sm}$ & $1.208$ & $\pm 0.054$ & $1.208$ & $\pm 0.054$ & $1.208$ & $\pm 0.054$ & $1.208$ & $\pm 0.054$ \\[0.5ex]
\hline
\rule{0 px}{15 px}
$y^\sm_e \;/\, 10^{-6}$ & $2.794745$ & $^{+0.000015}_{-0.000016}$ & $2.8482$ & $^{+0.0022}_{-0.0021}$ & $2.8646$ & $^{+0.0032}_{-0.0029}$ & $2.8782$ & $^{+0.0042}_{-0.0039}$ \\
\rule{0 px}{15 px}
$y^\sm_\mu \;/\, 10^{-4}$ & $5.899863$ & $^{+0.000019}_{-0.000018}$ & $6.0127$ & $^{+0.0047}_{-0.0044}$ & $6.0473$ & $^{+0.0067}_{-0.0062}$ & $6.0761$ & $^{+0.0088}_{-0.0082}$ \\
\rule{0 px}{15 px}
$y^\sm_\tau \;/\, 10^{-2}$ & $1.002950$ & $^{+0.000090}_{-0.000091}$ & $1.02213$ & $^{+0.00078}_{-0.00077}$ & $1.0280$ & $\pm 0.0011$ & $1.0329$ & $^{+0.0014}_{-0.0015}$ \\[0.5ex]
\hline
\rule{0 px}{15 px}
$g_3$ & $1.2143$ & $^{+0.0035}_{-0.0036}$ & $1.0560$ & $\pm 0.0023$ & $1.0017$ & $\pm 0.0020$ & $0.9510$ & $\pm 0.0017$ \\
\rule{0 px}{15 px}
$g_2$ & $0.65184$ & $^{+0.00018}_{-0.00017}$ & $0.63935$ & $\pm 0.00016$ & $0.63383$ & $\pm 0.00016$ & $0.62792$ & $\pm 0.00015$ \\
\rule{0 px}{15 px}
$g_1$ & $0.461425$ & $^{+0.000044}_{-0.000043}$ & $0.467774$ & $^{+0.000047}_{-0.000044}$ & $0.470767$ & $^{+0.000048}_{-0.000045}$ & $0.474110$ & $^{+0.000049}_{-0.000046}$ \\[0.5ex]
\hline
\end{tabular}
\caption{Values of the running SM quantities in the $\MSbar$ scheme at selected renormalisation scales $\mu$ and their marginalised highest posterior density (HPD) intervals (corresponding to the $1\sigma$ uncertainties). The gauge coupling $g_1$ is given in SU(5) normalisation $g_1^2 = 5/3 g'^2$. The uncertainties are calculated with a Monte Carlo analysis from the uncertainties for the input parameters as given in section 2. } \label{tab:SMvalues}
\end{table}

\begin{table}
\begin{center}
\setlength{\tabcolsep}{0.3em}
\begin{tabular}{|c|ll|ll|ll|}
\hline
\rule{0 px}{11 px}
SM Quantity & \multicolumn{2}{|c|}{$\mu = 1$ TeV} & \multicolumn{2}{|c|}{$= 3$ TeV} & \multicolumn{2}{|c|}{$= 10$ TeV} \\
\hline
\hline
\rule{0 px}{13 px}
$y^\sm_u \, /\, 10^{-6}$ & $6.3$ & $^{+1.3}_{-2.6}$ & $6.0$ & $^{+1.2}_{-2.4}$ & $5.6$ & $^{+1.2}_{-2.3}$\\
\rule{0 px}{15 px}
$y^\sm_d \, /\, 10^{-5}$ & $1.353$ & $^{+0.196}_{-0.087}$ & $1.282$ & $^{+0.189}_{-0.079}$ & $1.215$ & $^{+0.179}_{-0.075}$ \\
\rule{0 px}{15 px}
$y^\sm_s \, /\, 10^{-4}$ & $2.68$ & $^{+0.14}_{-0.15}$ & $2.54$ & $^{+0.13}_{-0.14}$ & $2.40$ & \
$^{+0.12}_{-0.13}$ \\
\rule{0 px}{15 px}
$y^\sm_c \,/\, 10^{-3}$ & $3.078$ & $\pm 0.094$ & $2.913$ & $^{+0.089}_{-0.090}$ & $2.758$ & $^{+0.084}_{-0.086}$ \\
\rule{0 px}{15 px}
$y^\sm_b \, /\, 10^{-2}$ & $1.376$ & $^{+0.013}_{-0.014}$ & $1.293$ & $\pm 0.013$ & $1.216$ & $^{+0.013}_{-0.012}$ \\
\rule{0 px}{15 px}
$y^\sm_t$ & $0.8613$ & $^{+0.0089}_{-0.0083}$ & $0.8216$ & $^{+0.0089}_{-0.0084}$ & $0.7841$ & $^{+0.0084}_{-0.0090}$ \\[0.5ex]
\hline
\rule{0 px}{15 px}
$\theta_{12}^{q,\sm}$ & $0.22736$ & $^{+0.00072}_{-0.00071}$ & $0.22736$ & $^{+0.00072}_{-0.00071}$ & $0.22736$ & $^{+0.00072}_{-0.00071}$ \\
\rule{0 px}{15 px}
$\theta_{23}^{q,\sm} \;/\, 10^{-2}$ & $4.296$ & $^{+0.066}_{-0.065}$ & $4.330$ & $\pm 0.066$ & $4.364$ & $^{+0.067}_{-0.066}$ \\
\rule{0 px}{15 px}
$\theta_{13}^{q,\sm} \;/\, 10^{-3}$ & $3.72$ & $\pm 0.13$ & $3.75$ & $\pm 0.13$ & $3.77$ & $^{+0.13}_{-0.14}$ \\
\rule{0 px}{15 px}
$\delta^{q,\sm}$ & $1.208$ & $\pm 0.054$ & $1.208$ & $\pm 0.054$ & $1.208$ & $\pm 0.054$ \\[0.5ex]
\hline
\rule{0 px}{15 px}
$y^\sm_e \, /\, 10^{-6}$ & $2.8501$ & $^{+0.0022}_{-0.0021}$ & $2.8664$ & $^{+0.0032}_{-0.0029}$ & $2.8800$ & $^{+0.0042}_{-0.0039}$ \\
\rule{0 px}{15 px}
$y^\sm_\mu \, /\, 10^{-4}$ & $6.0167$ & $^{+0.0047}_{-0.0044}$ & $6.0511$ & $^{+0.0067}_{-0.0062}$ & $6.0798$ & $^{+0.0088}_{-0.0082}$ \\
\rule{0 px}{15 px}
$y^\sm_\tau \, /\, 10^{-2}$ & $1.02281$ & $^{+0.00078}_{-0.00077}$ & $1.0287$ & $\pm 0.0011$ & $1.0335$ & $^{+0.0014}_{-0.0015}$ \\[0.5ex]
\hline
\rule{0 px}{15 px}
$g_3$ & $1.0597$ & $^{+0.0023}_{-0.0024}$ & $1.0049$ & $\pm 0.0020$ & $0.9538$ & $\pm 0.0017$ \\
\rule{0 px}{15 px}
$g_2$ & $0.63990$ & $\pm 0.00016$ & $0.63436$ & $\pm 0.00016$ & $0.62844$ & $^{+0.00015}_{-0.00016}$ \\
\rule{0 px}{15 px}
$g_1$ & $0.467774$ & $^{+0.000047}_{-0.000044}$ & $0.470767$ & $^{+0.000048}_{-0.000045}$ & $0.474110$ & $^{+0.000049}_{-0.000046}$ \\[0.5ex]
\hline
\end{tabular}
\end{center}
\caption{Values of the running SM quantities at $\mu = 1 \, \TeV, 3 \, \TeV$ and $10 \, \TeV$, converted to the $\DRbar$ scheme for use in an analysis above $M_\SUSY = \mu$, and their marginalised highest posterior density (HPD) intervals (corresponding to the $1\sigma$ uncertainties). The gauge coupling $g_1$ is given in SU(5) normalisation $g_1^2 = 5/3 g'^2$. The values can be used when applying the matching rules of Eqs.~(\ref{eq:matching_u_2}) - (\ref{eq:matching_e_2}) which include the $\tan\beta$-enhanced SUSY threshold corrections. The parameter $\bar\beta$ is defined in Eq.~(\ref{eq:def_barbeta}) such that it absorbs the threshold correction parameter for the tau lepton. If the threshold effects for the charged leptons are neglected, $\bar\beta$ simply reduces to the usual $\beta$. When calculating the uncertainties for the Yukawa couplings of the charged leptons, we recommend to use a relative uncertainty of 0.5\% instead of the smaller statistical error given above, in order to account for theoretical uncertainties, e.g.\ from the non-$\tan \beta$-enhanced SUSY threshold corrections.}
\label{tab:SMDRbarvalues}
\end{table}

\begin{figure}
\centering
          
        \begin{subfigure}[b]{0.450\textwidth}
                \centering
                \includegraphics[width=\textwidth]{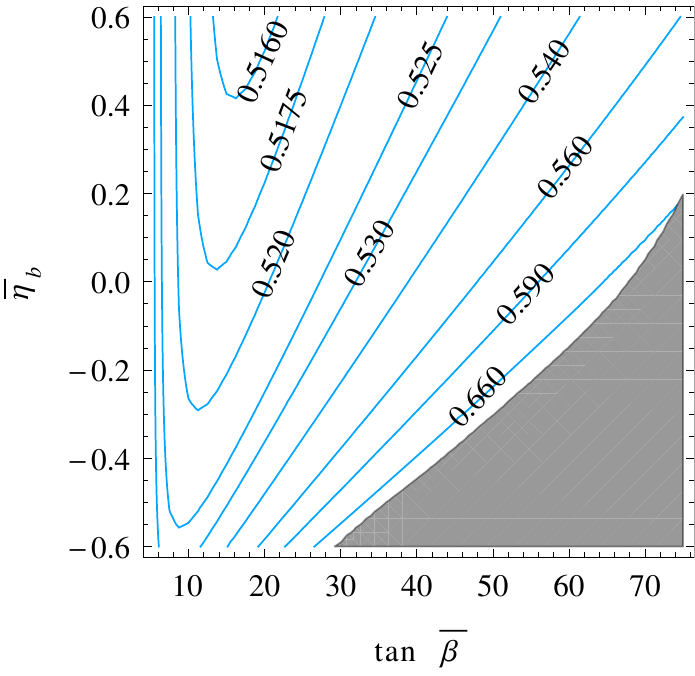}
                \vspace{-.5cm}
                \caption{$y^\mssm_t \; \sin\bar{\beta}$}
                \vspace{.5cm}
        \end{subfigure}
        \quad
        \begin{subfigure}[b]{0.450\textwidth}
                \centering
                \includegraphics[width=\textwidth]{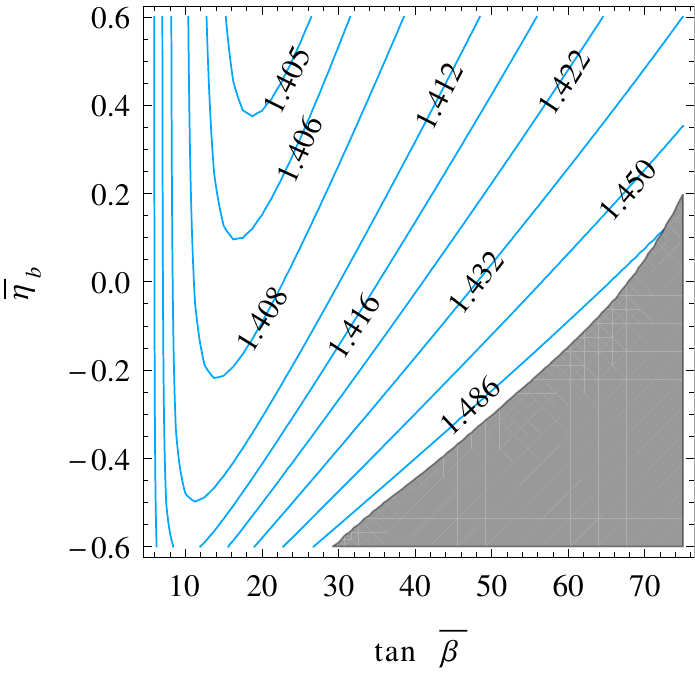}
                \vspace{-.5cm}
                \caption{$y^\mssm_c \; \sin\bar{\beta} \; / \; 10^{-3}$}
                \vspace{.5cm}
        \end{subfigure}
        
        \begin{subfigure}[b]{0.450\textwidth}
                \centering
                \includegraphics[width=\textwidth]{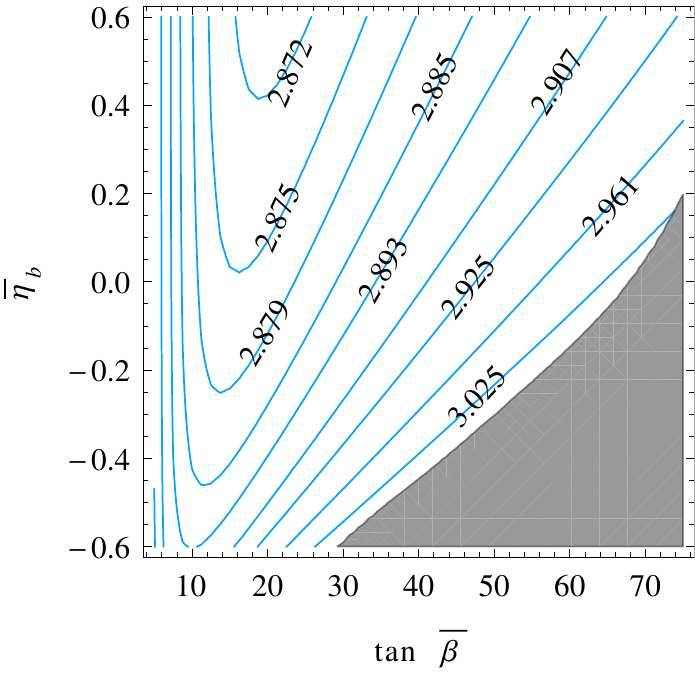}
                \vspace{-.5cm}
                \caption{$y^\mssm_u \; \sin\bar{\beta} \; / \; 10^{-6}$}
                \vspace{.5cm}
        \end{subfigure}
        
\caption{
Results for the running Yukawa couplings $y^\mssm_t$, $y^\mssm_c$ and $y^\mssm_u$ at $M_\GUT$ in the $\DRbar$ scheme, multiplied by $\sin\bar\beta$ (assuming $M_\SUSY = 1$ TeV). As discussed in section 3, the GUT scale quantities can be given to a good approximation as functions of only $\tan\bar\beta$ and $\bar{\eta}_b$. The parameters $\bar\beta$ and $\bar{\eta}_b$ are defined in Eqs.~(\ref{eq:def_barbeta}) and (\ref{eq:def_baretaB}), respectively. If the threshold effects for the charged leptons are neglected, $\tan \bar\beta$ simply reduces to the usual $\tan \beta$. In the dark grey regions of the plots, at least one of the Yukawa couplings becomes non-perturbative before $M_\GUT$. The relative uncertainties for the parameters are summarised in table \ref{tab:errors}. The dependence of the relative uncertainties for the third generation Yukawa couplings on $\tan\bar\beta$ and $\bar{\eta}_b$ is shown in figure \ref{fig:GUTscalevaluesUncertainty}. For the other parameters the relative uncertainties are  practically independent of $\tan\bar\beta$ and $\bar{\eta}_b$. For convenience, we also provide data tables for our GUT scale results at \url{http://particlesandcosmology.unibas.ch/RunningParameters.tar.gz}, which also contain the cases $M_\SUSY = 3$ TeV and $10$ TeV. From these tables one can easily obtain the GUT scale quantities as numerical functions of $\tan\bar\beta$ and $\bar{\eta}_b$ which can be conveniently used for model analyses. 
}
\label{fig:GUTscalevalues1}
\end{figure}

\begin{figure}
\centering
         
        \begin{subfigure}[b]{0.45\textwidth}
                \centering
                \includegraphics[width=\textwidth]{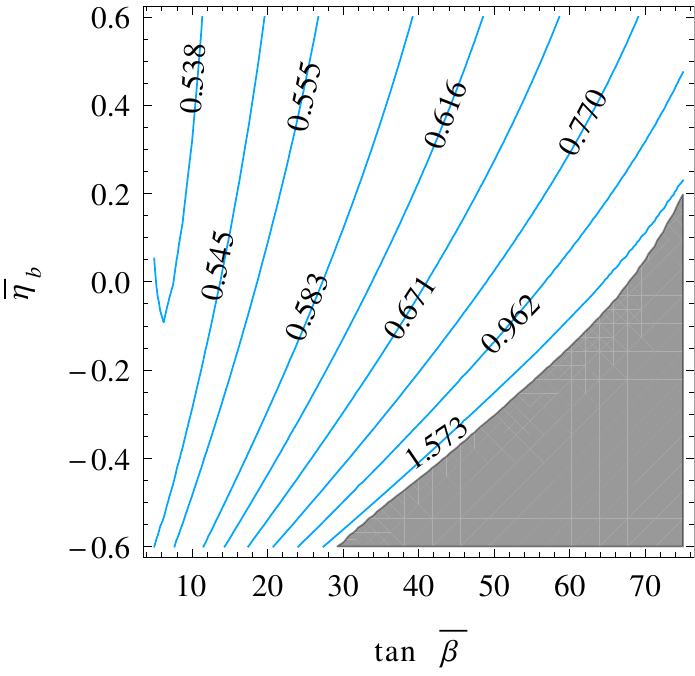}
                \vspace{-.5cm}
                \caption{$(1 + \bar{\eta}_b) \; y^\mssm_b \; \cos\bar{\beta} \; / \; 10^{-2}$}
                \vspace{.5cm}
        \end{subfigure}
        \quad      
        \begin{subfigure}[b]{0.45\textwidth}
                \centering
                \includegraphics[width=\textwidth]{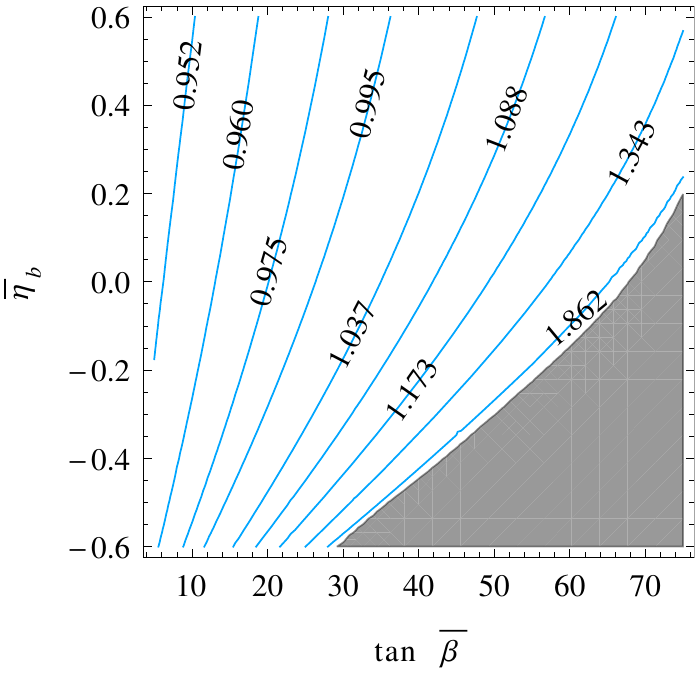}
                \vspace{-.5cm}
                \caption{$(1 + \bar{\eta}_q) \; y^\mssm_s \; \cos\bar{\beta} \; / \; 10^{-4}$}
                \vspace{.5cm}
        \end{subfigure}

        \begin{subfigure}[b]{0.45\textwidth}
                \centering
                \includegraphics[width=\textwidth]{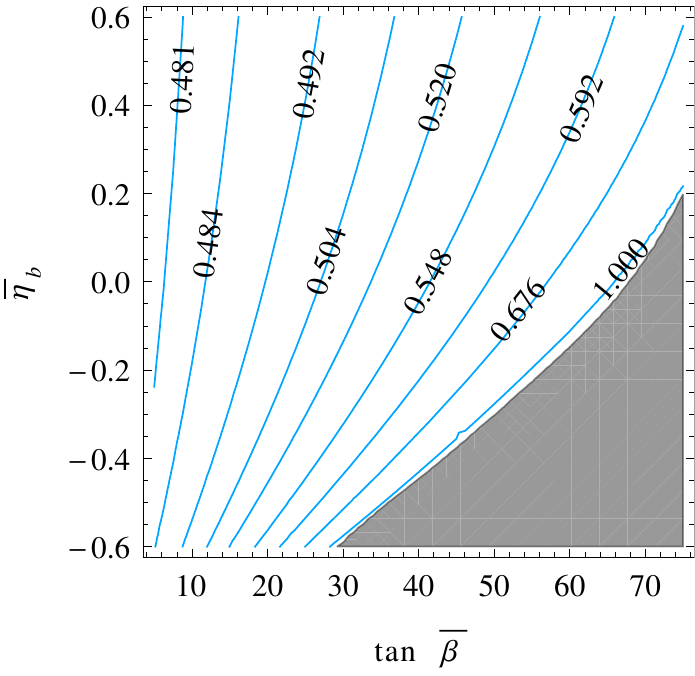}
               \vspace{-.5cm}
                \caption{$(1 + \bar{\eta}_q) \; y^\mssm_d \; \cos\bar{\beta} \; / \; 10^{-5}$}
                \vspace{.5cm}
        \end{subfigure}
 \caption{
Results for the running Yukawa couplings $y^\mssm_b$, $y^\mssm_s$ and $y^\mssm_d$ at $M_\GUT$ in the $\DRbar$ scheme (assuming $M_\SUSY = 1$ TeV), multiplied by the respective threshold correction factor and $\cos\bar\beta$ (cf.\ Eqs.~(\ref{eq:thres_yd}) and (\ref{eq:thres_yb})). As discussed in section 3, the GUT scale quantities can be given to a good approximation as functions of only $\tan\bar\beta$ and $\bar{\eta}_b$. The parameters $\bar\beta$ and $\bar{\eta}_b$ are defined in Eqs.~(\ref{eq:def_barbeta}) and (\ref{eq:def_baretaB}), respectively. If the threshold effects for the charged leptons are neglected, $\tan \bar\beta$ simply reduces to the usual $\tan \beta$. In the dark grey regions of the plots, at least one of the Yukawa couplings becomes non-perturbative before $M_\GUT$.  Further explanations can be found in the caption of figure \ref{fig:GUTscalevalues1}.}
\label{fig:GUTscalevalues2}
\end{figure}

\begin{figure}
\centering

        \begin{subfigure}[b]{0.45\textwidth}
                \centering
                \includegraphics[width=\textwidth]{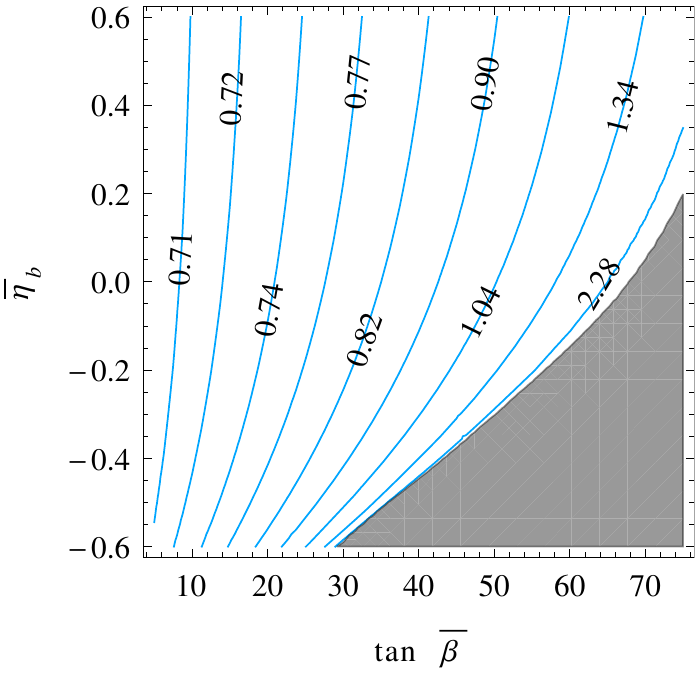}
                \vspace{-.5cm}
                \caption{$y^\mssm_\tau \; \cos\bar{\beta} \; / \; 10^{-2} $}
                \vspace{.5cm}
        \end{subfigure}
    \quad       
        \begin{subfigure}[b]{0.45\textwidth}
                \centering
                \includegraphics[width=\textwidth]{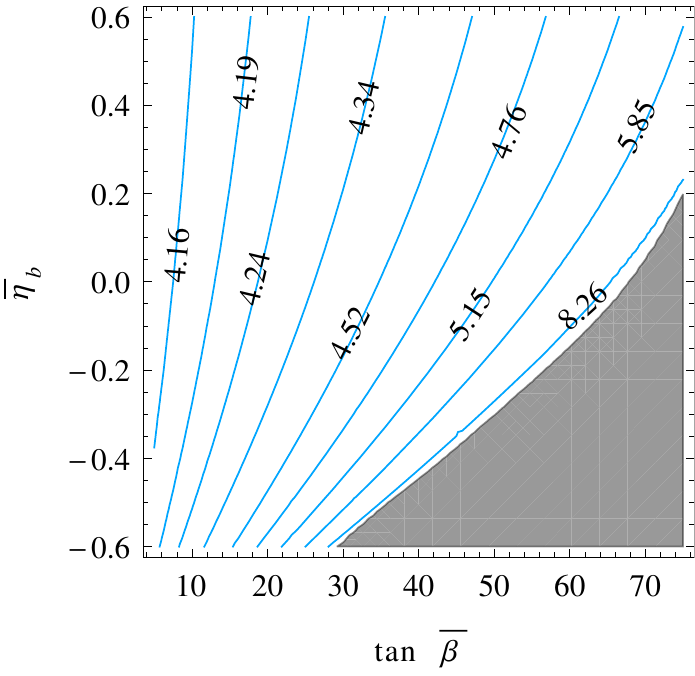}
               \vspace{-.5cm}
                \caption{$(1+\bar\eta_\ell) \; y^\mssm_\mu \; \cos\bar{\beta} \; / \; 10^{-4}$}
                \vspace{.5cm}
        \end{subfigure}

        \begin{subfigure}[b]{0.45\textwidth}
                \centering
                \includegraphics[width=\textwidth]{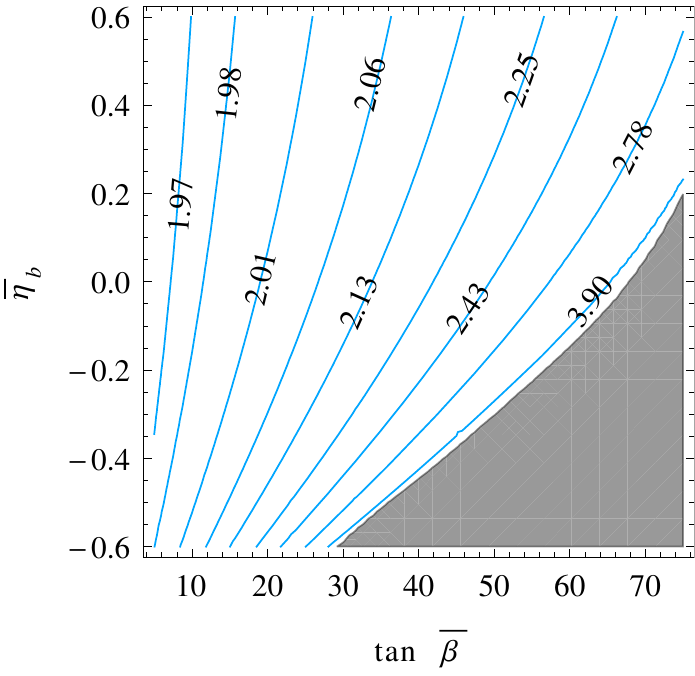}
               \vspace{-.5cm}
                \caption{$ (1+\bar\eta_\ell) \; y^\mssm_e \cos\bar{\beta} \; / \; 10^{-6} $}
                \vspace{.5cm}
        \end{subfigure}
\caption{
Results for the running Yukawa couplings $y^\mssm_\tau$, $y^\mssm_\mu$ and $y^\mssm_e$ at $M_\GUT$ in the $\DRbar$ scheme (assuming $M_\SUSY = 1$ TeV), multiplied by the respective threshold correction factor and $\cos\bar\beta$ (cf.\ Eqs.~(\ref{eq:thres_ye}) and (\ref{eq:thres_ytau})). As discussed in section 3, the GUT scale quantities can be given to a good approximation as functions of only $\tan\bar\beta$ and $\bar{\eta}_b$. The parameters $\bar\beta$ and $\bar{\eta}_b$ are defined in Eqs.~(\ref{eq:def_barbeta}) and (\ref{eq:def_baretaB}), respectively. If the threshold effects for the charged leptons are neglected, $\tan \bar\beta$ simply reduces to the usual $\tan \beta$. In the dark grey regions of the plots, at least one of the Yukawa couplings becomes non-perturbative before $M_\GUT$.  Further explanations can be found in the caption of figure \ref{fig:GUTscalevalues1}.
 }
\label{fig:GUTscalevalues3}
\end{figure}

\begin{figure}
\centering
        \begin{subfigure}[b]{0.450\textwidth}
                \centering
                \includegraphics[width=\textwidth]{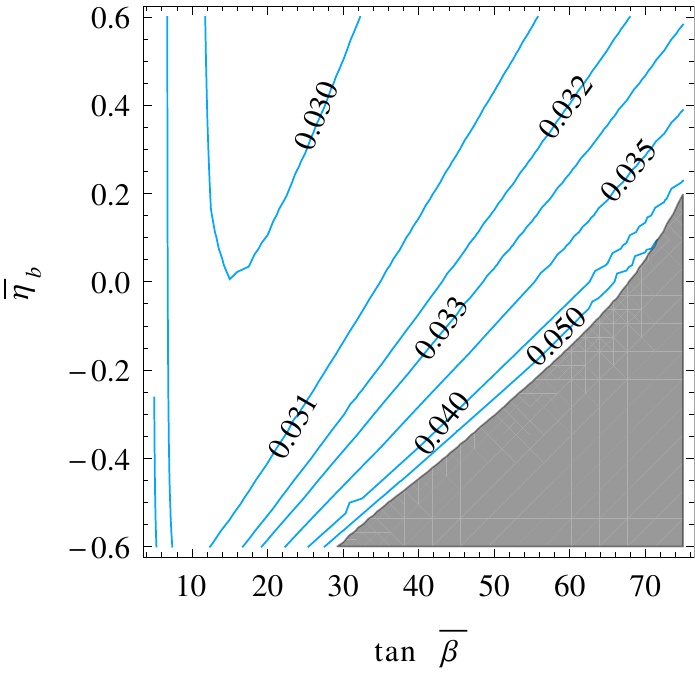}
                 \vspace{-.5cm}
                \caption{$\sigma(y^\mssm_t) / y^\mssm_t$}
                  \vspace{.5cm}
        \end{subfigure}
        \quad
        \begin{subfigure}[b]{0.450\textwidth}
                \centering
                \includegraphics[width=\textwidth]{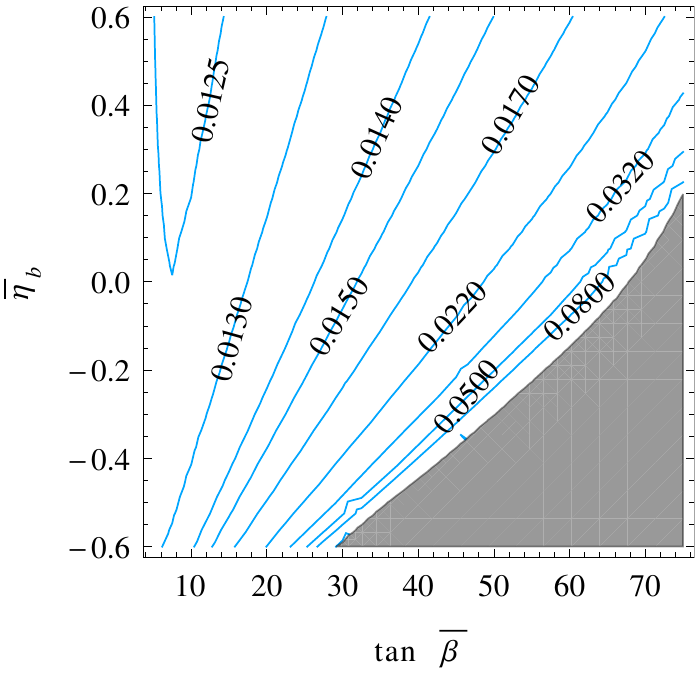}
                \vspace{-.5cm}
                \caption{$\sigma(y^\mssm_b) / y^\mssm_b$}
                 \vspace{.5cm}
        \end{subfigure}
         
        \begin{subfigure}[b]{0.450\textwidth}
                \centering
                \includegraphics[width=\textwidth]{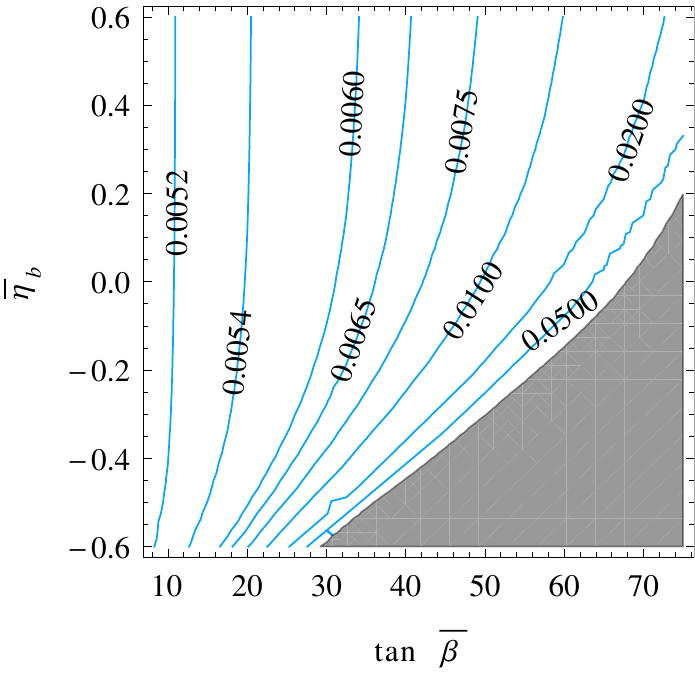}
                 \vspace{-.5cm}
                \caption{$\sigma(y^\mssm_\tau) / y^\mssm_\tau$}
                  \vspace{.5cm}
        \end{subfigure}
\caption{
Relative uncertainties for the GUT scale values of the running third generation Yukawa couplings $y^\mssm_t$, $y^\mssm_b$ and $y^\mssm_\tau$ (assuming $M_\SUSY = 1$ TeV). 
As for the GUT scale Yukawa couplings themselves, also their uncertainties can be given to a good approximation as functions of only $\tan\bar\beta$ and $\bar{\eta}_b$. The parameters $\bar\beta$ and $\bar{\eta}_b$ are defined in Eqs.~(\ref{eq:def_barbeta}) and (\ref{eq:def_baretaB}), respectively. If the threshold effects for the charged leptons are neglected, $\tan \bar\beta$ simply reduces to the usual $\tan \beta$. In the dark grey regions of the plots, at least one of the Yukawa couplings becomes non-perturbative before $M_\GUT$.  Further explanations can be found in the caption of figure \ref{fig:GUTscalevalues1}. The relative errors for the other parameters are summarised in table \ref{tab:errors}.
}
\label{fig:GUTscalevaluesUncertainty}
\end{figure}

\begin{figure}
\centering
        \begin{subfigure}[b]{0.450\textwidth}
                \centering
                \includegraphics[width=\textwidth]{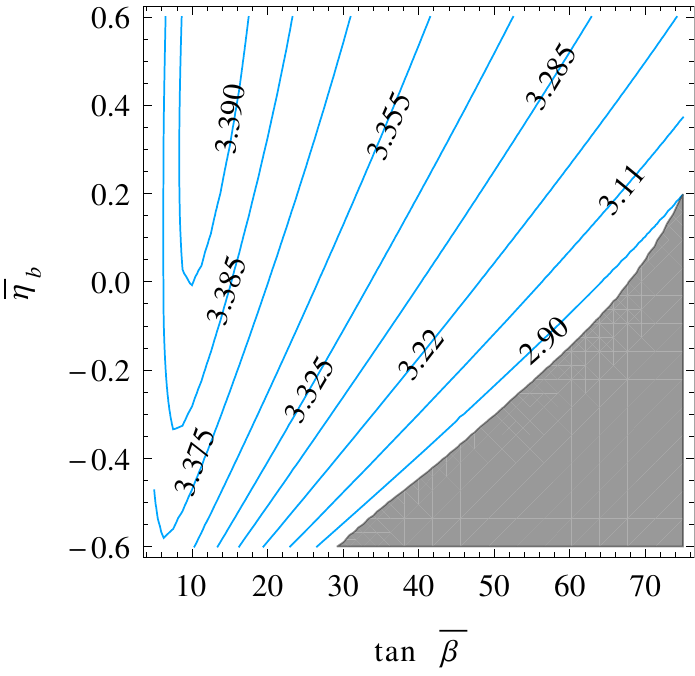}
                \vspace{-.5cm}
                \caption{$(1 + \bar{\eta}_q) / (1 + \bar{\eta}_b) \; \theta^{q,\mssm}_{13} \; / \; 10^{-3}$}
                \vspace{.5cm}
        \end{subfigure}
        \begin{subfigure}[b]{0.450\textwidth}
                \centering
                \includegraphics[width=\textwidth]{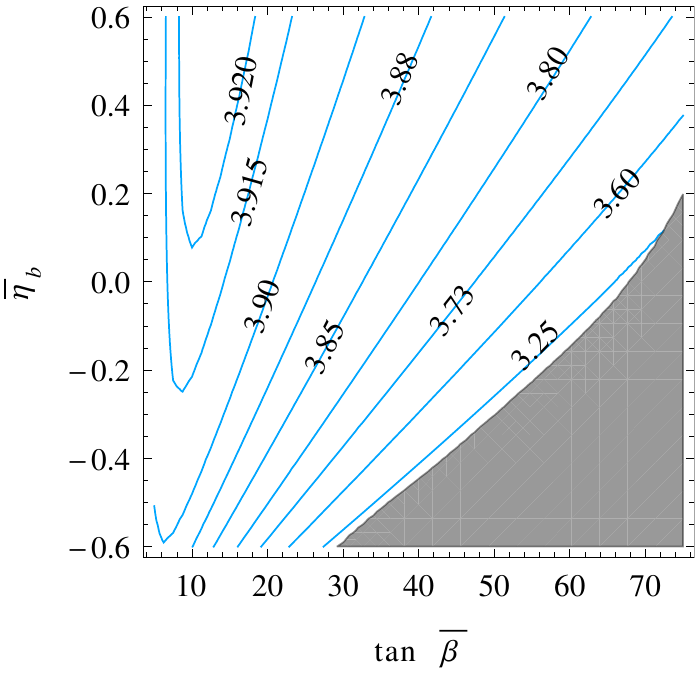}
                \vspace{-.5cm}
                \caption{$(1 + \bar{\eta}_q) / (1 + \bar{\eta}_b) \; \theta^{q,\mssm}_{23} \; / \; 10^{-2}$}
                \vspace{.5cm}
        \end{subfigure}

\caption{
Results for the CKM mixing angles $\theta_{13}^{q,\mssm}$ and  $\theta_{13}^{q,\mssm}$ at $M_\GUT$ in the $\DRbar$ scheme (assuming $M_\SUSY = 1$ TeV), multiplied by the respective threshold correction factor (cf.\ Eq.~(\ref{eq:thres_ti3})). 
The results of the mixing angles at the GUT scale can be obtained by multiplying the values in the plots by $[(1 + \bar{\eta}_q) / (1 + \bar{\eta}_b)]^{-1}$.
As discussed in section 3, the GUT scale quantities can be given to a good approximation as functions of only $\tan\bar\beta$ and $\bar{\eta}_b$. The parameters $\bar\beta$ and $\bar{\eta}_b$ are defined in Eqs.~(\ref{eq:def_barbeta}) and (\ref{eq:def_baretaB}), respectively. If the threshold effects for the charged leptons are neglected, $\tan \bar\beta$ simply reduces to the usual $\tan \beta$. In the dark grey regions of the plots, at least one of the Yukawa couplings becomes non-perturbative before $M_\GUT$.
}
\label{fig:GUTscalevalues4}
\end{figure}

\begin{table}
\begin{center}
\begin{tabular}{|c|c|}
\hline
\rule{0 px}{11 px}
MSSM Quantity $z_i$& Relative uncertainty ${\sigma(z_i) }/{ z_i}$ at $M_\GUT$ \\
\hline
\hline
\rule{0 px}{13 px}
$y^\mssm_u $ & 31\% \\
\rule{0 px}{15 px}
$y^\mssm_d $ & 11\% \\
\rule{0 px}{15 px}
$y^\mssm_s $ & 5.4\% \\
\rule{0 px}{15 px}
$y^\mssm_c$ & 3.5\% \\
\rule{0 px}{15 px}
$y^\mssm_b $ & see figure \ref{fig:GUTscalevaluesUncertainty} \\
\rule{0 px}{15 px}
$y^\mssm_t$ & see figure \ref{fig:GUTscalevaluesUncertainty} \\[0.5ex]
\hline
\rule{0 px}{15 px}
$\theta_{12}^{q,\mssm}$ & 0.32\% \\
\rule{0 px}{15 px}
$\theta_{23}^{q,\mssm}$ & 1.6\% \\
\rule{0 px}{15 px}
$\theta_{13}^{q,\mssm}$ & 3.6\% \\
\rule{0 px}{15 px}
$\delta^{q,\mssm}$ & 4.5\% \\[0.5ex]
\hline
\rule{0 px}{15 px}
$y^\mssm_e$ & 0.6\% \\
\rule{0 px}{15 px}
$y^\mssm_\mu$ & 0.6\% \\
\rule{0 px}{15 px}
$y^\mssm_\tau $ & see figure \ref{fig:GUTscalevaluesUncertainty} \\[0.5ex]
\hline
\end{tabular}
\end{center}
\caption{
Relative uncertainties $\sigma(z_i) / z_i$ for the running GUT scale parameters $z_i$ assuming $M_\SUSY = 1$ TeV. For the third generation Yukawa couplings the relative uncertainties depend on $\tan\bar\beta$ and $\bar{\eta}_b$ as shown in figure \ref{fig:GUTscalevaluesUncertainty}. 
As explained in the caption of table 2, we have used $0.5$\% relative uncertainty for the charged lepton Yukawa couplings at $M_\SUSY$ (which get enlarged to $0.6$\% uncertainty at $M_\GUT$, as shown in the table, due to RG effects and the uncertainties for the other parameters).
}
\label{tab:errors}
\end{table}

\clearpage

\begin{figure}
\centering
        \begin{subfigure}[b]{0.450\textwidth}
                \centering
                \includegraphics[width=\textwidth]{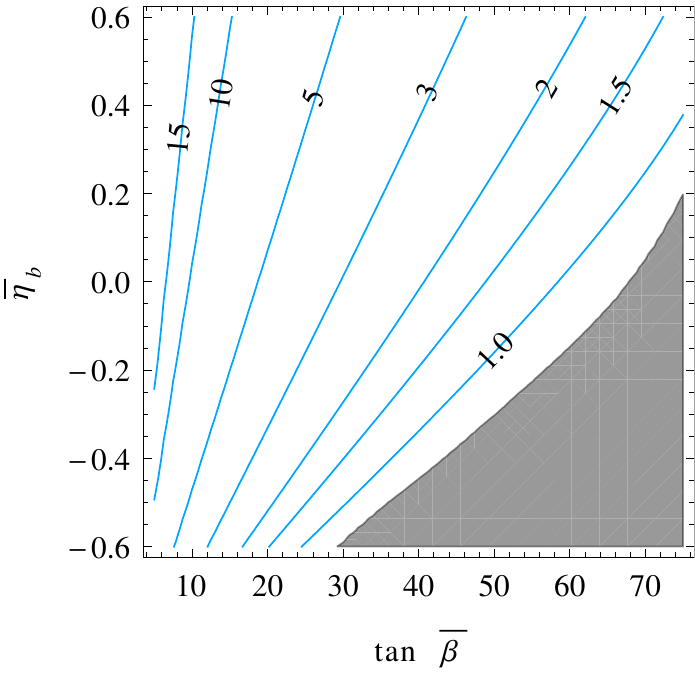}
                 \vspace{-.5cm}
                \caption{$y^\mssm_t/y^\mssm_b$}
                  \vspace{.5cm}
        \end{subfigure}
        \quad
        \begin{subfigure}[b]{0.450\textwidth}
                \centering
                \includegraphics[width=\textwidth]{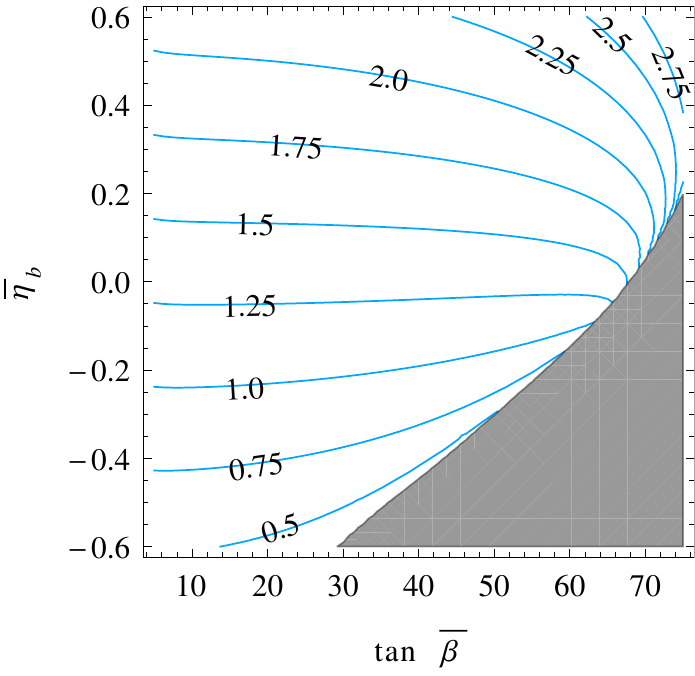}
                 \vspace{-.5cm}
                \caption{$y^\mssm_\tau/y^\mssm_b$}
                  \vspace{.5cm}
        \end{subfigure}
\caption{
Values of the GUT scale ratios $y^\mssm_t/y^\mssm_b$ and $y^\mssm_\tau/y^\mssm_b$ as functions of $\tan\bar\beta$ and $\bar{\eta}_b$ for $M_\SUSY = 1$ TeV. The parameters $\bar\beta$ and $\bar{\eta}_b$ are defined in Eqs.~(\ref{eq:def_barbeta}) and (\ref{eq:def_baretaB}), respectively. If the threshold effects for the charged leptons are neglected, $\tan \bar\beta$ simply reduces to the usual $\tan \beta$. In the dark grey regions of the plots, at least one of the Yukawa couplings becomes non-perturbative before $M_\GUT$.  
}
\label{fig:GUTscalevaluesRatios}
\end{figure}

\end{document}